# Microfluidic switchboards with integrated inertial pumps

Brandon S. Hayes[1,2] · Alexander N. Govyadinov[1] · Pavel E. Kornilovitch[1,3]

**Abstract** Arrays of H-shape microfluidic channels connecting two different fluidic reservoirs have been built with silicon/SU8 microfabrication technologies utilized in production of thermal inkjet printheads. The fluids are delivered to the channels via slots etched through the silicon wafer. Every H-shape channel comprises four thermal inkjet resistors, one in each of the four legs. The resistors vaporize water and generate drive bubbles that pump the fluids from the bulk reservoirs into and out of the channels. By varying relative frequencies of the four pumps, input fluids can be routed to any part of the network in any proportion. Several fluidic operations including dilution, mixing, dynamic valving, and routing have been demonstrated. Thus, a fully integrated microfluidic switchboard that does not require external sources of mechanical power has been achieved. A matrix formalism to describe flow in complex switchboards has been developed and tested.

## 1 Introduction

Thermal Inkjet (TIJ) printing was one of the earliest applications of MEMS and microfluidics (MF). Its commercial success fueled intense R&D work on TIJ-compatible materials, low-cost packaging, fluid handling methods, and robust fabrication processes. It also led to the establishment of high-volume MF production facilities around the world. All this knowledge base and infrastructure are now available for manufacturing MF components and modules for other applications such as chemical syntheses or biomedical analyses. To enable those, however, the MF toolbox needs to be expanded. At its core, TIJ printing is a simple one-fluid operation: a drop of ink is ejected from a nozzle by a vapor drive bubble and then the nozzle is refilled by capillary forces from an ink container [16]. There is no need for any sophisticated fluid routing or on-chip mixing: dots of four primary colors will mix on paper or other substrate producing a final image. For many non-printing applications, such simplicity is insufficient. For example, a typical biomedical application involves handling of several fluids (samples, reagents, buffers) in a series of process steps that may include filtering, mixing, heating, cooling, separation [11,6,13,5,14,9,1]. Thus, the potential of the TIJ infrastructure cannot be realized until the technology is shown to be extendable to more complex operations.

We previously reported how a basic TIJ component the TIJ resistor (a resistive microheater that vaporises a thin layer of fluid to produce a drive bubble) can be turned into a compact MF pump [19,10,8]. Because of their small dimensions ($< 1000$ $\mu$m$^2$) and compatibility with standard fab processes, such *inertial pumps* can be fabricated by hundreds or even thousands per chip with wide flexibility of sizes and placement locations. More generally, the TIJ resistors can be thought of as generic transducers of electrical to mechanical power which can drive MF systems in the same way power vias drive integrated electrical circuits.

In this paper, we report another extension of the fundamental TIJ technology. We fabricated H-shaped MF channels connected to two bulk fluid reservoirs.

[1]HP Inc., Corvallis, Oregon 97330, USA
[2]Rochester Institute of Technology, Rochester, NY, USA
[3]Physics Department, Oregon State University, Corvallis, OR, USA
E-mail: alexander.govyadinov@hp.com

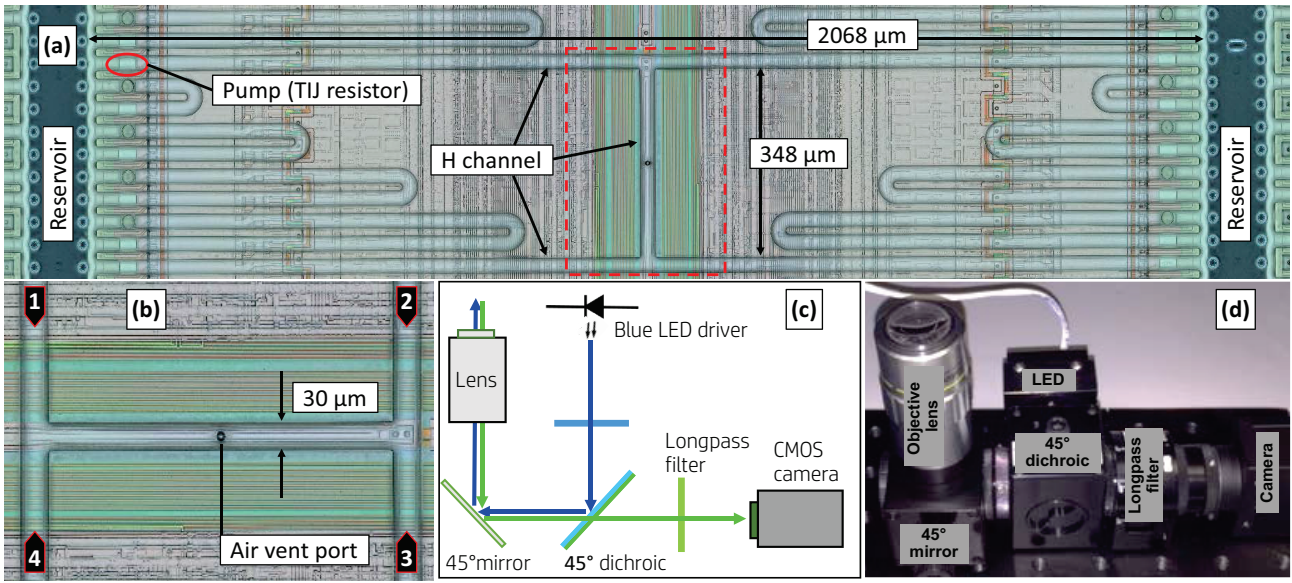

**Fig. 1** Fluidic design and optical setup. **a** Overall view of an H-shaped channel connecting two bulk fluid reservoirs. What is shown is 1/44th of an entire MF die that extends in the direction parallel to the reservoir slots. Light colored rectangles near the channels' ends are TIJ resistors that serve as inertial pumps. One of them is marked by a red oval. The U-shaped channels (five on each side of the image) are unrelated to the experiments described in this paper. **b** A magnified view of the area within a red dashed rectangle of panel **a**. The image is rotated 90° to correspond to the field of view of the optical system. The small circle in the center of the middle section is an air vent port in the channel "ceiling." Also shown is the fluidic ports' numbering convention (1, 2, 3, and 4) to be used later in the paper. **c** Optical diagram. **d** Photograph of the optical system (color figure online)

The H channel is a simple, but versatile MF device which was shown to work, for example, as a diffusion filter [2,3,12]. In this work, each leg of the H is equipped with its own inertial pump. The pulse frequency of the pumps can be individually adjusted in a quasi-continuous fashion. By varying the frequencies, several two-fluid and three-fluid operations including mixing, dilution, and routing are achieved. We also experimentally demonstrate the use of inertial pumps as dynamic valves, a concept described previously [7]. More generally, an H channel with pumps becomes a 4-terminal fluidic switchboard; depending on the mode of operation, each leg can serve as either fluidic input or output and each input fluid can be routed to any output at any proportion and flow rate desired.

## 2 Design and fabrication

The overall design is shown in Fig. 1a. TIJ resistors and fluidic layers were fabricated on top of 8-inch Si wafers that contained FET transistors and other drive electronics. The resistors were $20 \times 30$ $\mu$m$^2$ in size and arranged in 4 columns than run along the length of $25 \times 5$ mm$^2$ die. The column pitch was 42 $\mu$m. The H channels were positioned between the two inner columns. Physical dimensions are indicated in the figure. The length of each external leg was 1019 $\mu$m and of the central section

348 $\mu$m. The channel width was 30 $\mu$m, making the total reservoir-to-reservoir distance $2 \times 1019 + 30 = 2068$ $\mu$m.

The fluidic layer was built by using a two-step SU8 process [16]. First, the channel geometry was defined by photolithography and then a 20 $\mu$m SU8 dry film was laminated on top of the entire wafer providing "ceiling" for the channels. The channel width and height were 30 $\mu$m and 31 $\mu$m, respectively, so the channel cross-section was approximately square. Next, slots were etched through the silicon opening access to fluid reservoirs to be attached to the back of the die. Each die had two slots, thus providing the desired two-fluid functionality. Finally, the wafers were diced; individual die were assembled into printheads by gluing them to ceramic holders and fluid reservoirs with flex circuits attached for electrical connection. In total, each die contained 44 individual H-shaped channels and each wafer had 204 die.

Figure 1b shows a magnified view of the central section of the H, rotated 90° to match the views produced by the optical system. Also shown is the numbering convention (1 through 4) utilized in the rest of the paper to distinguish four external ports of the switchboard.

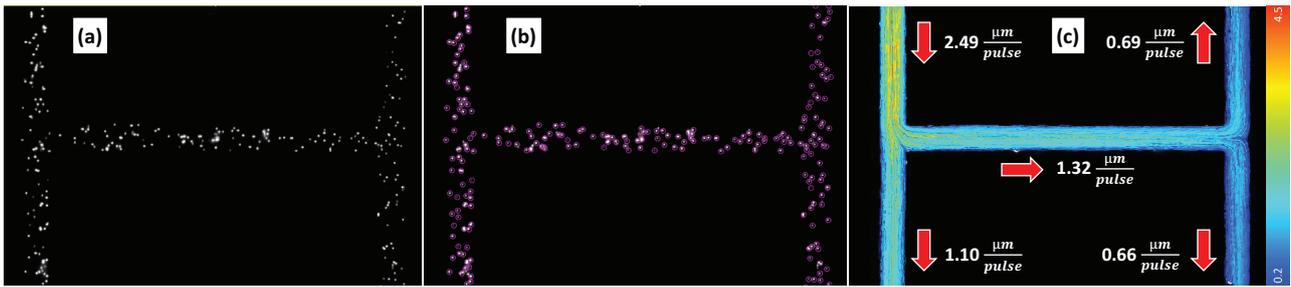

**Fig. 2** Video processing sequence. The top left pump is pulsed at 13.33 Hz. **a** Snapshot of an original video. **b** Same frame with tracking circles assigned to each fluorescent particle. **c** A resulting flow field color coded by average velocity. Linear velocity in $\mu$m/pulse can be converted to $\mu$m/s by multiplying by the pump frequency. It can also be converted to flow rates in pL/pulse by multiplying with the conversion factor $30 \cdot 31/1000 = 0.93$ pL/$\mu$m, where $30 \times 31$ $\mu$m$^2$ is the cross-sectional area of the channel. Note conservation of mass at both T-junctions (color figure online)

## 3 Experiment

### 3.1 Fluid priming and pump operation

In all experiments, deionized water with surfactant (Dow Corning 67 Additive) was utilized. To visualize the flow, Polyscience Inc. Fluoresbrite® YG Carboxylate microspheres 1.75 $\mu$m in diameter were added. Original stock of 2.5 wt.% was diluted from 25 to 500 times depending on the concentration required for reliable visualization and tracking. In order to prevent evaporation and drooling through nozzles and vent ports, a thin film of SU8 was applied to the die surface. SU8 was attached to the die surface by heating the film for initial bonding to the SU8 channels and then UV exposure to cross-link the polymer. The fluids were pipetted into the bulk reservoirs after which they passively wicked through the pathways to fill the H channels. Care had to be taken to not trap air in the system – especially in the central section of the H. Hydrostatic pressure on two sides of the H had to be equalized by establishing a macroscopic fluidic connection between the two bulk reservoirs away from the channels; otherwise, the pressure difference caused a constant flow through the side legs of the H. The printheads were operated in a reservoir-up/die-down orientation to prevent the fluids from dripping out.

TIJ technology allows varying the Si die operating temperature between 28°C and 85°C with ±1°C accuracy [19]. In the present work, temperature was set at (28±1)°C. However, *local* fluid temperature can change in response to pump frequency, resulting in some variation of fluid viscosity. To select an operating firing energy, duration of electrical firing impulses was varied from 0.9 to 1.5 $\mu$s with total energy per electrical impulse in the range of $0.8 - 1.4$ $\mu$J. The vapor drive bubble was maximal at 1.25 $\mu$J, which determined operating firing conditions. The TIJ electronic driver enables firing frequencies from 0 to 48 kHz. In the present work, short 10 kHz firing sequences followed by appropriate time delays were utilized, which resulted in time average frequencies of under 200 Hz.

### 3.2 Optical setup

The optical diagram is shown in Fig. 1c and real setup in Fig. 1d. Observations were performed using a custom designed inverted optical microscope with magnification from 10x to 200x. The basic design described earlier [19] was modified to enable fluorescent imaging capability by adding a set of filters and dichroic beam splitter using illumination from Lightspeed Technologies HPLS-36DD18B series single emitter with a built in driver for strobing a 478 nm blue LED. Fluorescent imaging allowed detection of 1.75 $\mu$m microspheres used to sample the flow. The camera frame capture was set to occur at the onset of each electrical impulse sequence. A quiet time between sequences of 75 ms was utilized for camera exposure. This ensured that the most amount of light would be collected by the camera between the firing events. An exposure greater than the physical quiet time would result in smearing of particles in the image. With current settings, a 75 ms period corresponds to a frame rate of 13.33 Hz. In order to maximize the perceived intensity of the fluorescent particles, the camera's gain was adjusted. After obtaining sufficient contrast between the fluorescent particles and the background, the firing conditions were set and videos longer than 1000 frames were recorded. Large video files were needed in order to generate enough statistics to fully characterize the flow profiles in the channels.

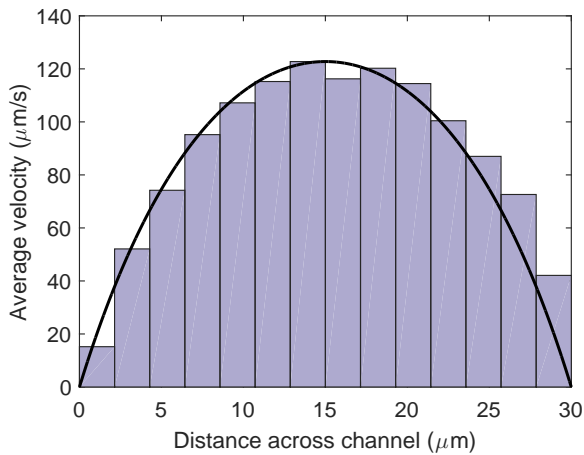

**Fig. 3** A typical experimental velocity histogram of flow through port 2 overlaid on the $z$-averaged theoretical velocity profile [18,20,4] in a $30 \times 31$ $\mu m^2$ rectangular channel with the same flow rate

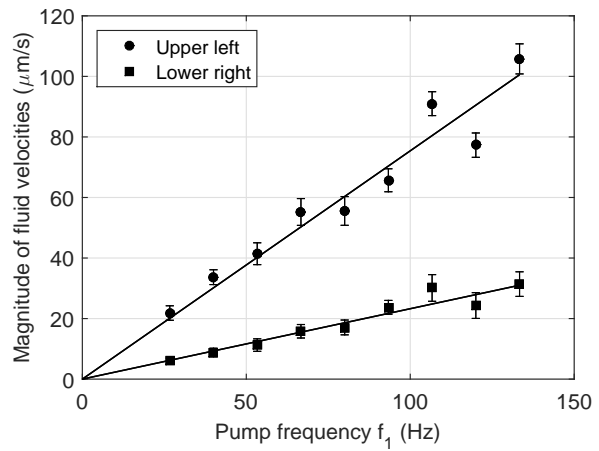

**Fig. 4** Fluid velocities in two legs of an H vs. top left pump frequency $f_1$ with other pumps idle, cf. Fig. 2. Linear dependencies are expected. The straight lines are guides for the eye

3.3 Particle tracking and post-processing

Videos were post-processed using Fiji [15], a distribution of ImageJ. Figure 2 illustrates the progression of the tracking process at each step. See Supplementary Information (videos S1 and S2) for full visualization of the tracking process. The Trackmate plugin [17] in Fiji was utilized to perform Lagrangian particle tracking of 1.75 $\mu$m particles in the videos — exemplified in Fig. 2a. Shown in Fig. 2b, a LoG (Laplacian of Gaussian) detection method was utilized to detect particles in the image while a linear motion LAP (Linear Assignment Problem) tracker was utilized to track particle movement in the video. Figure 2c visualizes the tracking process by coloring each particle track according to its calibrated average velocity in $\mu$m/pulse. As such, the particle tracks represent flow streamlines through the H. After tracking, the data was exported to .xml files for later processing and analysis. In order to achieve accurate tracking results, the initial search radius of the linear motion LAP tracker must be set to the maximum displacement of the particles. Therefore, fast moving particles generate larger displacements between frames which requires a larger initial search radius. However, as the initial search radius approaches the mean separation distance between particles, tracking errors will increase. Thus, the particle density must be maintained low enough in order to prevent such tracking error, but high enough in order to pass a sufficient number of particles to obtain the flow profile, which necessitates long video recordings.

After tracking, videos were post-processed using custom MATLAB scripts. These scripts consisted of 5 main parts: (1) file directory parser, (2) manually/automatically detect edges of channels to subdivide H into the five regions (upper right, lower right, upper left, lower left, and middle), (3) track particles in each part of H, (4) record magnitude of velocity of each particle and location, and (5) visualize results. Videos were spatially calibrated to the 30 $\mu$m channel width observed in each video. Histograms of average velocity across the channel width were generated to verify uniform particle distributions. Figure 3 highlights experimental agreement with the classical 2D rectangular flow profile [18, 4] collapsed along one dimension. Symmetry and flow rate continuity were checked for each video to verify accuracy of the tracking method. In addition, at sub kHz operating frequencies, flow induced by successive pulses does not overlap in time and linear scaling of flow rates with pump frequency is expected. (When the pulses overlap, pump efficiency per pulse goes down, but not significantly [8].) The linear dependence was verified experimentally as shown in Fig. 4.

4 Pump calculus

Before presenting results specific to H switchboards, it is useful to discuss the general principles of organizing flows in fluidic switchboards driven by inertial pumps. Consider a general N-port network illustrated in Fig. 5. Every electrical firing impulse applied to a TIJ resistor creates a high-pressure vapor bubble that generates pulse-like flow through the network [8]. For our channel dimensions, the flow lasts for approximately 100 $\mu$s. If the time interval between the impulses exceeds this characteristic time, the flow pulses are *independent* in the sense that the total volume of fluid moved as a re-sult of a pulse sequence is equal to a sum of volumes displaced in individual events. Such a switchboard is said to be *linear*. Nonlinear switchboards with interaction between pulses is an important but complex area. The single-pump nonlinearity was studied in [8] and found to be relatively weak. In contrast, interaction between different pumps is not weak and can actually be quite significant. As an example, two neigbouring pumps pulsing simultaneously can produce a flow that is quite different from sequential pulses. This topic is outside of the scope of the present study and is left for future work.

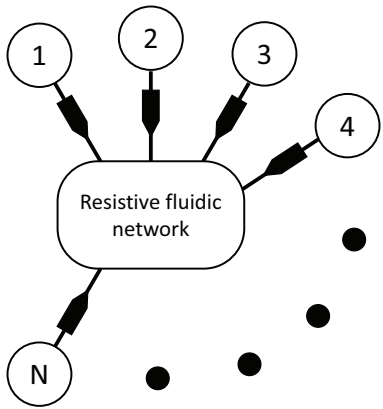

**Fig. 5** An abstract fluidic switchboard with $N$ ports. The black arrows indicate positive flow directions

We further assume that individual pulse characteristics are stable in time. Each pump generates a series of identical pulses as long as the fluid properties, such as temperature or viscosity, remain steady. Experience with thermal inkjet shows that pulse-to-pulse variation for the same resistor is less than 1%, which can be neglected for the purposes of this paper. Pump-to-pump variations across a wafer due to design differences and fabrication imperfections are typically less than 4%. Under the above assumptions, each switchboard is characterized by a set of numbers $V_{nm}$ defined as the volume of fluid passing through the $n$th port after a pulse generated by $m$th pump. $V_{nm}$ is dependent on the pump placement, firing impulse parameters, fluid properties, and the fluidic resistance of the channels in the network. By convention, each port is assigned a flow direction that is positive when the flow is from the reservoir toward the switchboard and negative when the flow is from the switchboard into the reservoir. Typically, $V_{mm} > 0$ and $V_{n \neq m, m} < 0$. That is, each pump pumps fluid from its respective reservoir into the other $(N-1)$ reservoirs. Note that due to conservation of fluid volume, $\sum_{n=1}^{N} V_{nm} = 0$ for all $m$. It is assumed that the central part of the switchboard does not contain sources or sinks or accumulators of fluid. Now let each pump operate at frequency $f_m$. Then, flow rate through the $n$th port, measured in [volume/second], is given by

$$Q_n = \sum_{m=1}^{M} V_{nm} \cdot f_m \,. \quad (1)$$

$Q_n$ comprises contributions from its "own" $n$th pump that provides a positive flow rate and all other pumps that provide a negative flow rate. Depending on the ratio of pump frequencies, each port can become either fluidic input or output. The flow rates satisfy a global sum rule, $\sum_{n=1}^{N} Q_n = 0$, which also follows from volume conservation. If $V_{nm}$ are known, Eq. (1) can be regarded as a general relation between pump frequencies and flow rates. In the forward problem, $f_m$ are set and then, Eq. (1) predicts expected flow rates.

Consider now the inverse problem when desired rates $Q_n$ are specified and $f_m$ are adjusted to satisfy (1). The solution seems to be straightforward: just solve the set of linear equations (1) for $f_m$ for fixed $Q_n$. However, the procedure is ill-defined. Due to volume conservation at every pump pulse (all columns in $V$ sum to zero), one of the equations in (1) is a consequence of the other $(N-1)$ ones. Hence, the determinant of $V$ is zero and the matrix cannot be inverted. It means that one of the pump frequencies is a free parameter that has to be fixed by an additional condition. The inevitable presence of one free parameter in passive switchboards can be easily understood for highly symmetric fluidic networks like the H-channel considered in this work. Suppose that one particular set of frequencies $f'_m$ providing a desired flow pattern $Q_n$ is found. Since all pumps are of the same strengths, firing all four pumps in succession once will not change the overall balance: whatever volume is pulled from $m$th reservoir by firing $m$th pump, will be pushed back by firing the other three pumps. Continuing the argument, increasing (or decreasing) all $f'_m$ by the same increment $\Delta f$ does not change the overall flow rates as partial contributions from the pumps cancel each other. Hence, the frequencies are defined up to one free parameter.

Returning to inverting the system (1), one pump frequency, $f_m$, has to be set externally. Which pump to fix and to what value is a matter of choice and depends on switchboard details. After the choice is made, other $f_{m' \neq m}$ can be determined from $(N-1)$ equations of (1). It might happen that one or more frequencies $f_{m' \neq m}$ come out negative which is physically impossible. If that happens, one needs to set $f_m$ to a different value and repeat the process. In the end, all positive configurations $f_m$ consistent with the desired $Q_n$ can be identified.

We close this section by drawing an analogy between Eq. (1) and the theory of electric circuits. Consider a *linear* electric circuit with $N$ voltage sources. Set the first source to a unit voltage and the others to zero. Then write down and solve Kirchhoff's laws to determine currents flowing through the circuit branches. The currents will be proportional to the source voltage. Those relationships are analogous to the first column of Eq. (1), with the source voltage taking the role of pump frequency. By repeating the process for all voltage sources, an entire matrix is constructed. The electric current flowing through a particular branch is a superposition of currents induced by individual voltage sources. This is in perfect analogy with fluidic networks where the flow rate through a particular port is a superposition of elementary flows induced by individual inertial pumps, as expressed by Eq. (1).

## 5 Results

We now use the general principles to describe flows through the H-shape switchboards.

### 5.1 One-pump operations: pump matrix

As described in Sect. 4, firing each pump of the H in independent videos was utilized to construct its representative pump matrix $V$. Thus, each video of the $m$th pump firing generates a column $V_{:m}$. Mathematically, the theoretical pump matrix

$$V_{\text{th}} = \begin{pmatrix} (2a+b) & -a & -a & -b \\ -a & (2a+b) & -b & -a \\ -a & -b & (2a+b) & -a \\ -b & -a & -a & (2a+b) \end{pmatrix}, \quad (2)$$

is highly symmetric due to the H geometry and singular due to volume conservation. Consider the first column (pump position 1, cf. Fig. 1b): $V_{11}$ has flow coming from the reservoir to the channels and thus positive by convention. It consists of the outflow from the channels to reservoirs through ports 2 and 3, which are equal by symmetry, $V_{12} = V_{13} = -a$, plus the outflow $V_{14} = -b$ through port 4. This logic is repeated for all four pumps in the matrix. Thus, the theoretical pump matrix can be characterized by just two parameters: (1) the total moved volume per pulse $(2a+b)$ and (2) the split ratio at the first T-junction $b/(2a)$. In practice, manufacturing variations make $V$ slightly different for different H's and measurement errors introduce noise.

Equation (3) shows an experimental matrix of one H-shaped channel measured with the four-video procedure described above:

$$V = \begin{pmatrix} 2.26 & -0.73 & -0.52 & -1.02 \\ -0.65 & 2.58 & -1.23 & -0.67 \\ -0.63 & -1.19 & 2.63 & -0.68 \\ -1.06 & -0.85 & -0.62 & 2.54 \end{pmatrix} \text{[pL/pulse]}, \quad (3)$$

which is not fully symmetric because of noise. Volume conservation is satisfied reasonably well, as all the columns in $V$ sum up to zero with an average error of less than 7%.

*A priori* knowledge of network symmetries and conservation laws can be used to improve the experimental matrix to smooth over noise. In this approach, the four pump videos and subsequent analyses provide four independent measurements of the two matrix parameters, and the results are averaged to finalize the numbers. That leads to a corrected experimental pump matrix

$$V_{\text{corr}} = \begin{pmatrix} 2.49 & -0.68 & -0.68 & -1.13 \\ -0.68 & 2.49 & -1.13 & -0.68 \\ -0.68 & -1.13 & 2.49 & -0.68 \\ -1.13 & -0.68 & -0.68 & 2.49 \end{pmatrix} \text{[pL/pulse]}. \quad (4)$$

$V_{\text{corr}}$ was derived from $V$ using a two-step procedure. First, similar parameters [all $a$'s, all $b$'s, and all $(2a+b)$'s in Eq. (2)] were averaged in $V$ and replaced with the averaged result. Second, the uncertainties in $a$ and $b$ were assumed to contribute equally to the volume conservation error. Then both $a$ and $b$ were adjusted by the same absolute value $\Delta = 0.006$ to eliminate the error. Thus, the experimental matrix (3) was forced into the theoretical form of Eq. (2). It can now be applied to predict flow patterns for various frequency vectors $f_m$.

### 5.2 Two-pump operations: dynamic valve, controlled dilution

Consider a switchboard operating with pumps 1 and 4 active and the other two pumps idle. The flow pattern follows from Eq. (2) for $f_2 = f_3 = 0$. One obtains for the ratio of flows through ports 4 and 1

$$\frac{Q_4}{Q_1} = \frac{(2a+b) - b\frac{f_1}{f_4}}{(2a+b)\frac{f_1}{f_4} - b}. \quad (5)$$

There are two special points in this expression, both corresponding to conditions of *dynamic valving* [7]. At $f_1/f_4 = (2a+b)/b$, flow through port 4 is exactly zero. At this frequency ratio, pump 4 pumps just enough to push back any fluid sent through port 4 by pump 1, see Fig. 6a and ESI video S3. Thus, pump 4 effectively blocks flow through port 4 and acts as a dynamic valve. This type of valve may be *leaky*. While the average

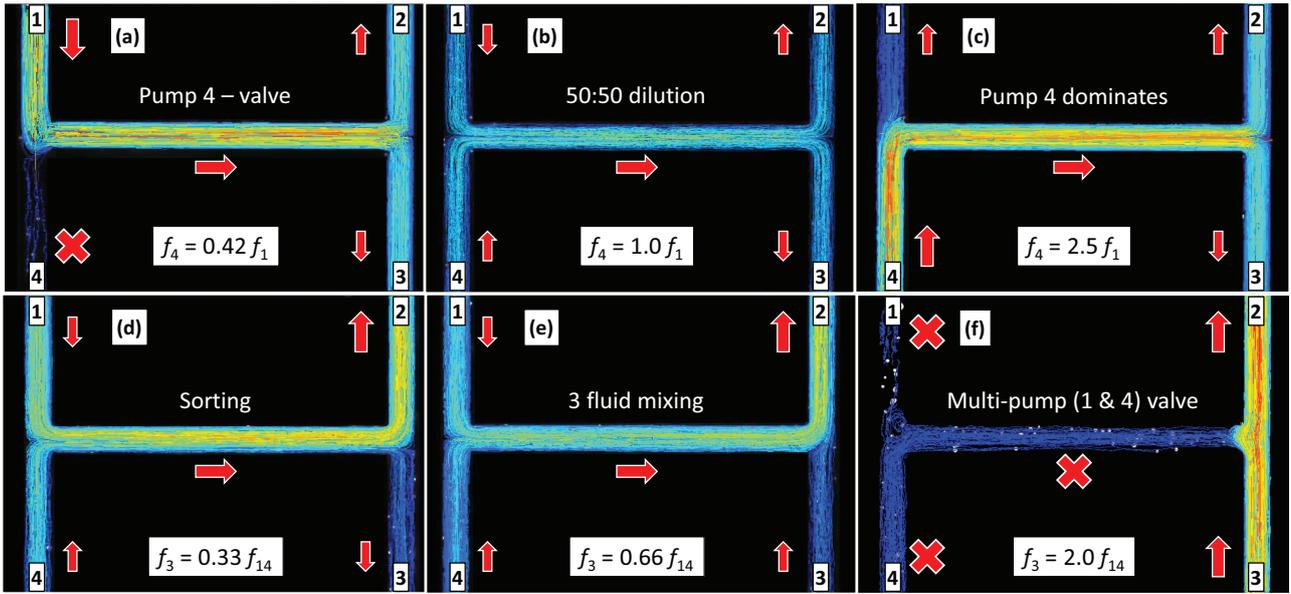

**Fig. 6** Flow regimes of an H switchboard. **a-c** Two-pump regimes with $f_2 = f_3 = 0$ and increasing ratio $f_4/f_1$. **a** Pump 4 pumps just enough to block flow through port 4. $f_4/f_1 = 1/2.4$. **b** Pumps 1 and 4 pump equal amounts into the middle section achieving $50:50$ dilution of two fluids. $f_4/f_1 = 1.0$. **c** Pump 4 overpowers pump 1. $f_4/f_1 = 2.5$. **d-f** Three-pump regimes with $f_2 = 0$, $f_1 = f_4 = f_{14} = 40$ Hz, and increasing ratio $f_3/f_{14}$. **d** Activation of pump 3 regulates the split between outputs 2 and 3, thereby routing the product flowing out of the H's middle section. $f_3/f_{14} = 1/3$. **e** As $f_3$ increases, flow through port 3 reverses direction. Inputs from ports 1, 3, and 4 flow into one output 2, achieving three-fluid mixing. $f_3/f_{14} = 2/3$. bf f At $f_3 = 2\,f_{14}$, flow in the middle section stops, creating a multi-pump valve (color figure online)

flow rate may be zero, a complex velocity profile across the channel cross section may lead to the fluid slowly creeping past the valve along channel walls, according to CFD simulations [7]. At the inverse frequency ratio $f_1/f_4 = b/(2a+b)$, flow through port 1 is zero, and pump 1 acts as a dynamic valve. Function (5) is plotted in Fig. 7.

The two valve points divide the entire $f_1/f_4$ interval into three regimes. At $f_1/f_4 > (2a+b)/b$, the flow pattern is dominated by pump 1 where fluid is drawn from port 1 and pushed into the other three ports. At $f_1/f_4 < b/(2a+b)$, the situation reverses and flow becomes dominated by pump 4, see Fig. 6c and ESI video S5. The most interesting regime lies between the two valving points. In this case, both pumps 1 and 4 provide positive flow rates through their respective ports, which means that the central section of H receives a mixture of fluids from ports 1 and 4, as shown in Fig. 6b and ESI video S4. Therefore, in a system where each port is connected to a different fluid reservoir, the two fluids are diluted at a ratio dependent on the firing frequency ratio.

In order to validate these theoretical concepts, a series of videos with different ratios $f_1/f_4$ were recorded and flow rates were measured using the particle tracking method described above. The results are shown in Fig. 7 overlaid on the theoretical curve of Eq. (5) for the pump

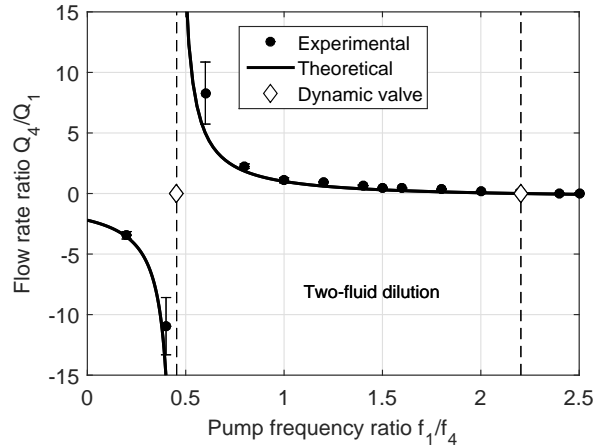

**Fig. 7** Two-pump switchboard operation. A two-fluid dilution regime lies between two dynamic valving points marked by diamonds. The solid line is the theoretical curve of Eq. (5) for the parameters of Eq. (4): $a = 0.68$ and $b = 1.13$. Negative values at $f_1/f_4 < 0.45$ mean that the flow direction in port 1 is negative (i.e., into reservoir 1)

matrix of Eq. (4). The agreement is within experimental errors. These results confirm that the matrix formalism developed here can predict dilution ratios of constituent liquids in complex fluidic switchboards. Note that Eq. (5) can be easily inverted to express $f_1/f_4$ via

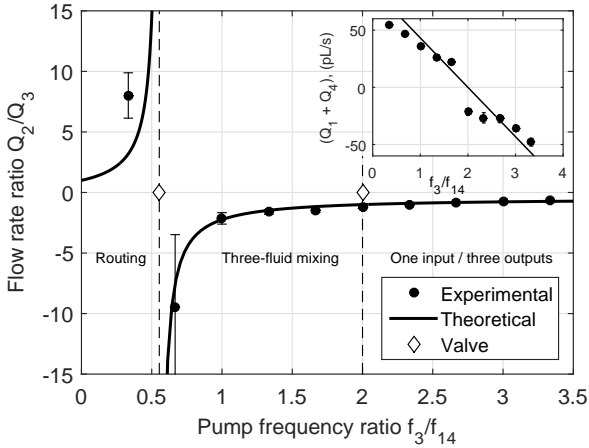

**Fig. 8** Three-pump switchboard operation for $f_1 = f_4 = 40$ Hz, $f_2 = 0$ and varying $f_3$. The two valving points, single pump at $f_3/f_{14} = 0.55$ and multi-pump at $f_3/f_{14} = 2.0$, divide the frequency interval into three regimes: routing, three-fluid mixing and "one input/three outputs." The solid line is Eq. (7) for $a = 0.54$ and $b = 0.87$. Inset: flow rate through the middle section of H in pL/s compared with the theoretical prediction of Eq. (8) (solid line, not a fit)

$Q_4/Q_1$. That is, if the target dilution ratio is known, the corresponding pump frequency ratio can be computed.

Note that pump 1 can also be paired up with another pump, for example, pump 2, to produce a different flow pattern and different dilution conditions. Flow specifics can be deduced from the same pump matrix. Any pair-wise pump operation in any fluidic network can be analyzed using the present method.

### 5.3 Three-pump operations. Routing. Three-fluid mixing

Three-pump operation provides the H switchboard with additional functionality. Assuming that some flow is generated by pumps 1 and 4 in the middle segment of the H, activation of pumps 2 and 3 controls split of the output flow between ports 2 and 3. Consider for simplicity the operating conditions when pumps 1 and 4 are pulsed at the same frequency $f_1 = f_4 = f_{14}$, pump 2 is idle, $f_2 = 0$, and $f_3$ is varied. Flow rate through port 3 follows from Eq. (2)

$$Q_3 = -2a\,f_{14} + (2a + b)\,f_3 \, . \quad (6)$$

Again, there is a dynamic valving point at $f_3 = 2a\,f_{14}/(2a+b) \equiv f_s$, when pump 3 completely blocks flow through port 3. A mixture of fluids 1 and 4 flowing through the central section of the H (to be called "product") is diverted to port 2. By symmetry, if pump 2 is operated instead and pump 3 is idle, the product will be diverted to port 3. Thus, at this frequency the switchboard operates as a *router*: the product is diverted to outputs 2 or 3 depending on which pump, 3 or 2, is activated. Note that the router can be converted into a *sorter* by adding an appropriate sensor to determine the state of a flowing product and a feedback loop to provide instructions where a particular portion of the product should be diverted to. At $f_3 < f_s$, the product is split between output ports 2 and 3 in proportion dependent on $f_3$:

$$\frac{Q_2}{Q_3} = \frac{2a + b\,\frac{f_3}{f_{14}}}{2a - (2a + b)\,\frac{f_3}{f_{14}}} \, . \quad (7)$$

An example of this regime is shown in Fig. 6d and ESI video S6. When $f_3 > f_s$, pump 3 draws additional fluid from port 3 and mixes it with the product flowing from the middle section. This is the case of *three-fluid mixing*: fluids from three different ports 1, 3, and 4 are pushed into one output port 2, see Fig. 6e and ESI video S7. Relative volume fractions in the mixture can be easily determined from the matrix formalism.

To validate the above concepts, another H-shaped channel from a different Si die was characterized and its pump matrix determined with the method of Sect. 5.1. It resulted in a symmetrized matrix of Eq. (2) with $a = 0.54$ pL/pulse and $b = 0.87$ pL/pulse. Then a series of videos for $f_{14} = 40$ Hz and different $f_3/f_{14}$ were recorded and processed using particle tracking. The experimental flow ratios shown in Fig. 8 agree well with the theoretical prediction of Eq. (7).

Upon further increase of $f_3$, another special point is encountered. At large frequencies, pump 3 pumps so much that it overcomes the combined action of pumps 1 and 2 and reverses flow in the middle section of the H. To see that, compute flow rate in the middle section as a sum of $Q_1$ and $Q_4$. It follows from Eq. (2) that

$$Q_1 + Q_4 = 2a\,(2f_{14} - f_3) \, , \quad (8)$$

where left-to-right is chosen as positive flow direction. Thus, at a critical ratio $f_3/f_{14} = 2.0$ flow stops. This is a *multi-pump dynamic valving* special point: combined action of pumps 1 and 4 is just enough to block flow produced by pump 3 in the central section, and the entire input from port 3 flows into port 2 and nowhere else, see Fig. 6f and ESI video S8. Notice that the critical frequency ratio 2.0 is independent of $a$ and $b$, which is a consequence of the H's high symmetry. At $f_3 > 2f_{14}$, fluid in the central section begins to flow right to left, and the switchboard enters a "one-input-three-outputs" regime. Experimental flows through the central section are shown in the inset of Fig. 8 and are in good agreement with Eq. (8).

We close this section by discussing the sources of variation in matrix parameters $a$ and $b$ observed in different experiments. One primary source is fabricational imperfections. Variations in the channels' width and height, inlet-outlet geometry, and active resistor area can result in device-to-device variations of fluidic resistances on the order of several percent. The inertial pump is a highly nonlinear device [10], which amplifies variations of displaced volume to tens of percent. Secondly, flow rates are very dependent on the physical properties of the fluid under test, primarily temperature, which affects the strength of vapour bubbles, and viscosity. Given that viscosity itself is a sharp function of temperature and that the local device temperature changes in response to pump frequency, the difference between two distinct experiments can reach hundreds of percent. A practical solution to this challenge would be robust and ubiquitous flow metering coupled with feedback loops to appropriately adjust pump frequencies and maintain the desired flow rates.

## 6 Summary and prospects

In the not so distant future, complex multifunctional microfluidic devices will be as ubiquitous as multifunctional electronic devices today. They will be an integral part of the physical-to-digital convergence and will power important applications such as chemical and biological syntheses, biomedical and environmental tests, and high-precision dispense of drugs and other valuable fluids. Knowledge base and infrastructure are already in place to manufacture complex microfluidic devices reliably and economically. What is lacking, however, is experience in large-scale simultaneous handling of multiple fluids and managing interactions between them.

The goal of this work was not only to demonstrate novel fluidic effects but to do so using scalable devices and processes used in mass production of inkjet printheads. Repurposing TIJ resistors as pumps to power fluidic networks is particularly attractive. Due to its small size ($< 1000$ $\mu$m$^2$) such pumps can be made in thousands per square centimeter and can provide unprecedented fluid handling functionalities. With this in mind, one of HP's commercial two-color printheads was modified to create a series of two-reservoir 4-port networks. Driven by TIJ inertial pumps, the networks became fluidic switchboards, where fluids could be routed from any input to any output at any rate by simply changing relative frequencies of the four pumps. Even a seemingly simple switchboard such as the H presented in this work possesses a rich variety of flow regimes. Single-pump and multi-pump dynamic valving, two-fluid and three-fluid dilution, and controlled routing have all been demonstrated. We plan on fabricating and characterizing more complex switchboards in the future.

One element needed to operate multi-port switchboards is a general theory that can predict flow rates. In this paper, such a theory has been formulated and tested on the H's. In a perfect analogy with electrical circuits, the pumps play the role of voltage sources (with voltage proportional to pump frequency), flow rates are analogous to electric currents, and reservoirs are grounds. The matrix formalism developed in Sect. 4 is nothing but a version of Kirchhoff's laws applied to microfluidics. Despite its simplicity, the theory has proved useful in classifying flow regimes of the H switchboard as detailed in Sect. 5.

**Acknowledgements** The authors wish to thank their HP Inc. colleagues: Ricky Brenneman, Galen Cook, Pirooz Fariborzi, Ed Friesen, Michael Hager, Susanne Muhly, and Nicholas McGuinness for device fabrication; David Markel and Erik Torniainen for fluid dynamics modelling; Daniel Curthoys, Tom Deskins, and Brian Morrissette for help with measurements; Michael Polander and Masoud Zavarehi for help with particle tracking algorithms; Michele Briggs, Daniel Gardner, Diane Hammerstad, Cheryl Macleod, Katrina Sloma, and Ken Vandehey for general support of this work.

# Supplementary videos for Figure 2

**Fig. 2a (S1)**

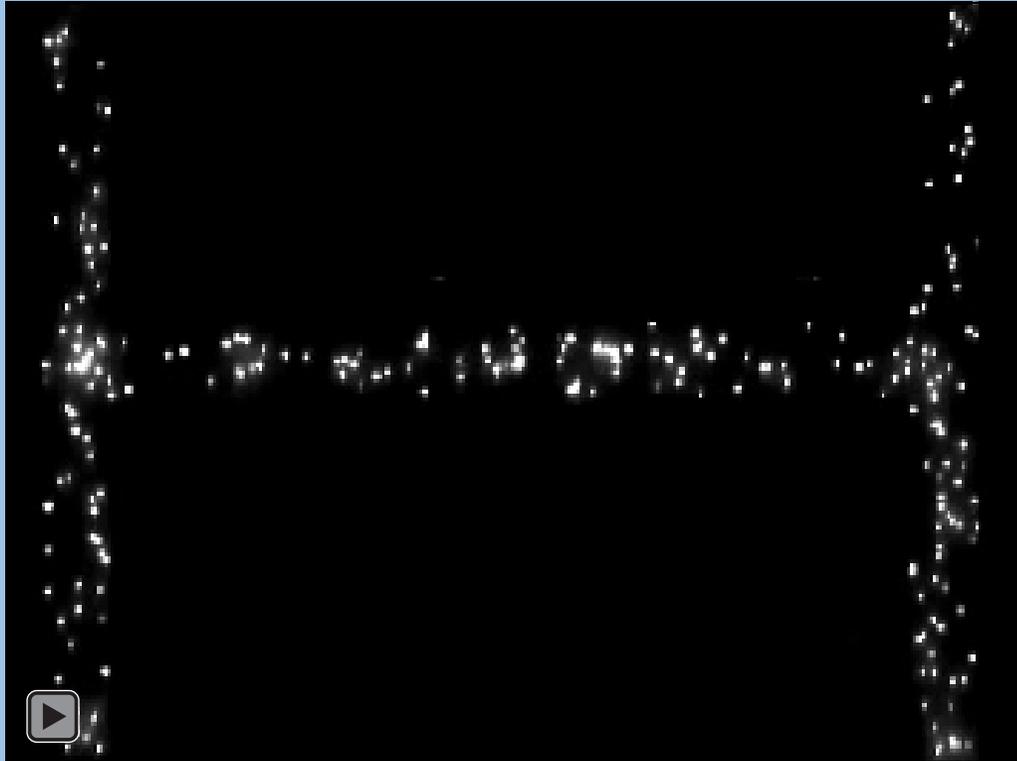

**Fig. 2b (S2)**

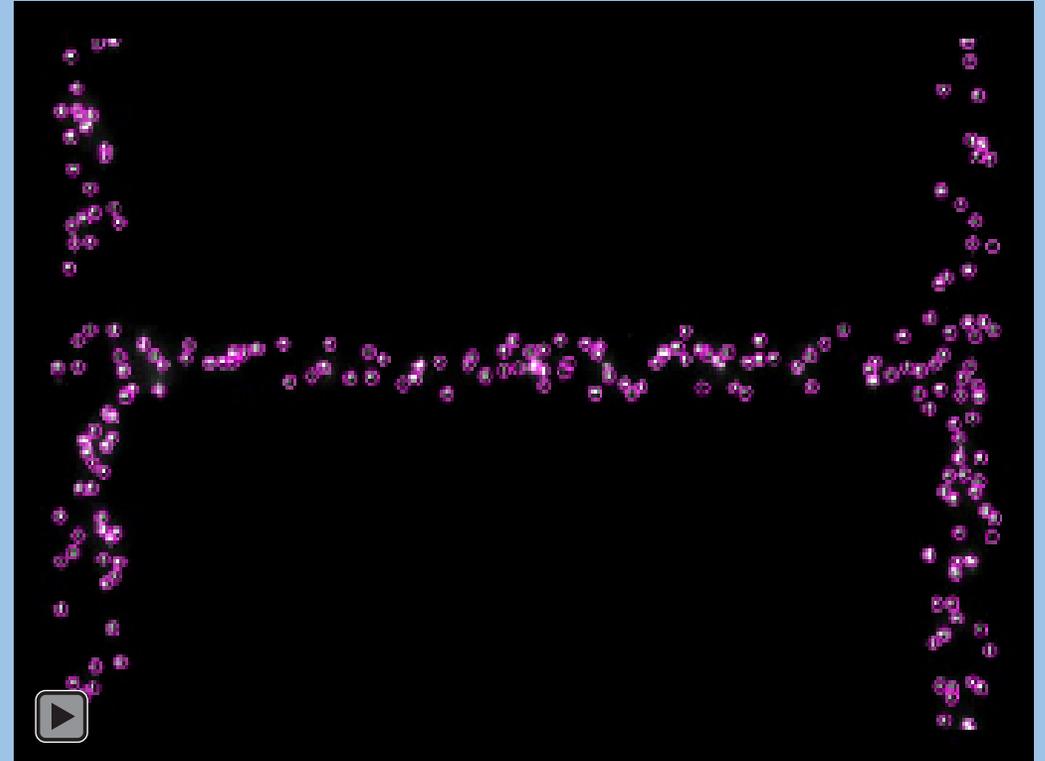

# Supplementary videos for Figure 6

### Fig. 6a (S3)
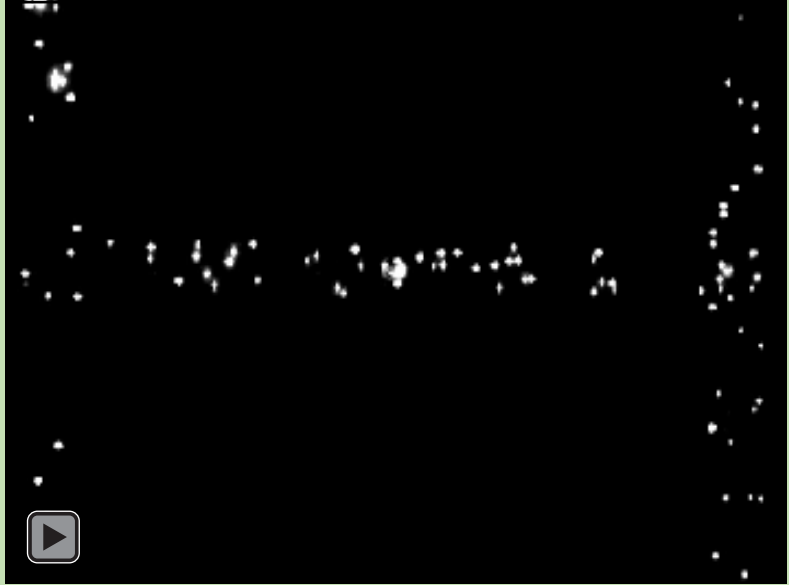

### Fig. 6b (S4)
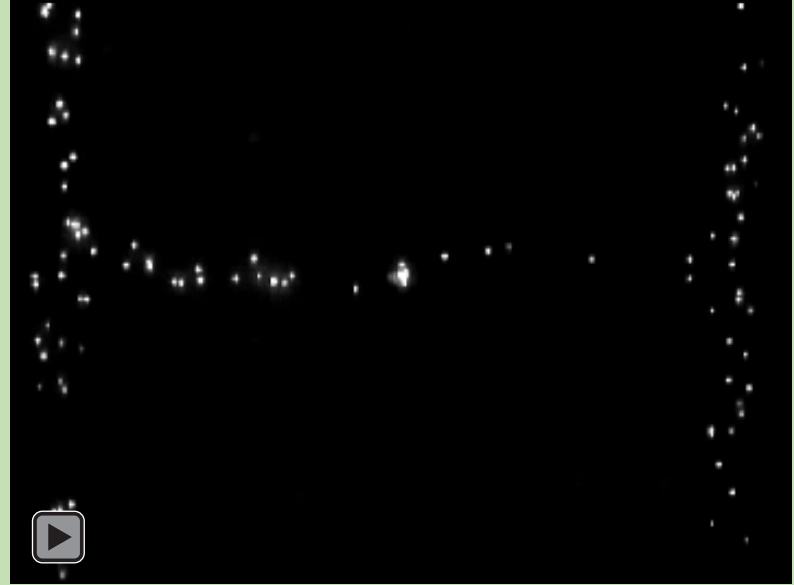

### Fig. 6c (S5)
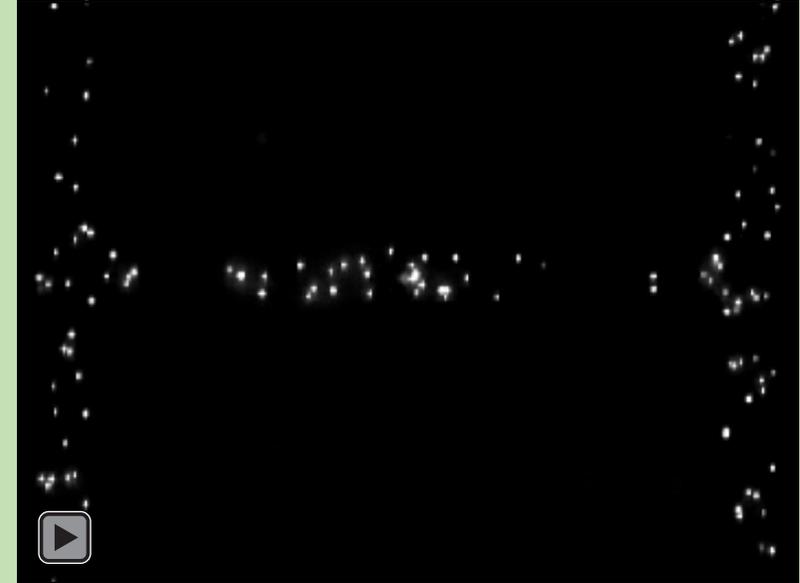

### Fig. 6d (S6)
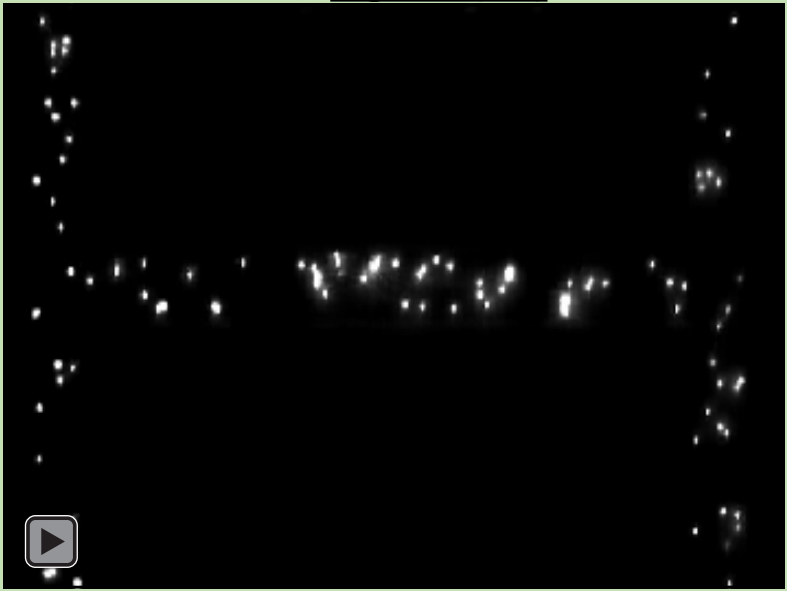

### Fig. 6e (S7)
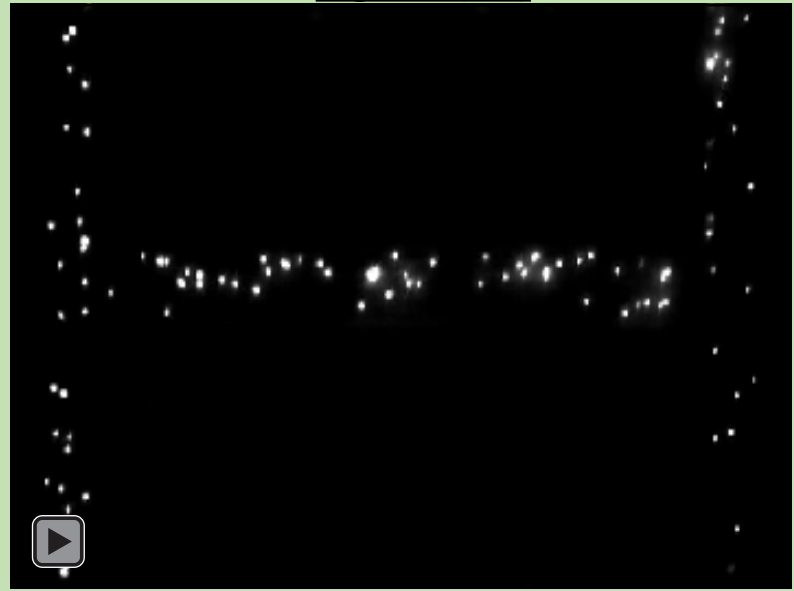

### Fig. 6f (S8)
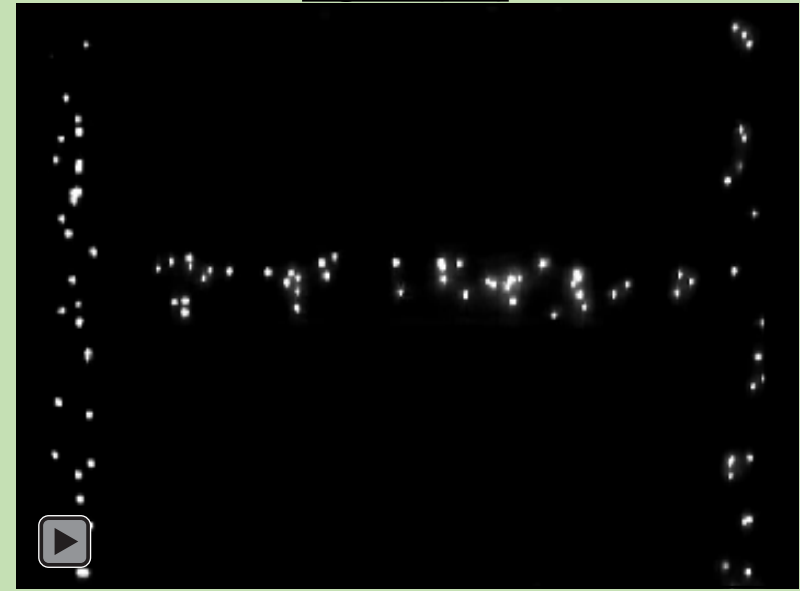